\documentstyle[preprint,prc,aps,floats,epsf]{revtex}

\tightenlines

\begin{document}

%
%
\def\symdef#1#2{\def#1{#2}}

\symdef{\alphabar}{\overline\alpha}
\symdef{\alphabarp}{\overline\alpha\,{}'}
\symdef{\Azero}{A_0}

\symdef{\bzero}{b_0}

\symdef{\Cs}{C_{\rm s}}
\symdef{\cs}{c_{\rm s}}
\symdef{\Cv}{C_{\rm v}}
\symdef{\cv}{c_{\rm v}}

\symdef{\dalem}{\frame{\phantom{\rule{8pt}{8pt}}}}
\symdef{\del}{\partial}
\symdef{\delN}{{\wt\partial}}
\symdef{\Deltaevac}{\Delta{\cal E}_{\rm vac}}

\symdef{\ed}{{\cal E}}
\symdef{\edk}{{\cal E}_k}
\symdef{\edkzero}{{\cal E}_{k0}}
\symdef{\edv}{{\cal E}_{\rm v}}
\symdef{\edvphi}{{\cal E}_{{\rm v}\Phi}}
\symdef{\edvphizero}{{\cal E}_{{\rm v}\Phi 0}}
\symdef{\edzero}{{\cal E}_{0}}
\symdef{\Efermistar}{E_{{\scriptscriptstyle \rm F}}^\ast}
\symdef{\Efermistarzero}{E_{{\scriptscriptstyle \rm F}0}^\ast}
\symdef{\etabar}{\overline\eta}
\symdef{\ezero}{e_0}

\symdef{\fomega}{f_\omegav}
\symdef{\fpi}{f_\pi}
\symdef{\fv}{f_{\rm v}}
\symdef{\fvt}{\widetilde\fv}

\symdef{\gA}{g_A}
\symdef{\gammazero}{\gamma_0}
\symdef{\gomega}{g_\omegav}
\symdef{\gpi}{g_\pi}
\symdef{\grho}{g_\rho}
\symdef{\grad}{{\bbox{\nabla}}}
\symdef{\gs}{g_{\rm s}}
\symdef{\gv}{g_{\rm v}}

\symdef{\fm}{\mbox{\,fm}}

\symdef{\infm}{\mbox{\,fm$^{-1}$}}
\symdef{\isovectorTensor}{s_{\tauvec}}
\symdef{\isovectorTensorN}{\wt\isovectorTensor}
\symdef{\isovectorVector}{j_{\tauvec}}
\symdef{\isovectorVectorN}{\wt j_{\tauvec}}

\symdef{\kappabar}{\overline\kappa}
\symdef{\kfermi}{k_{{\scriptscriptstyle \rm F}}}
\symdef{\kfermizero}{k_{{\scriptscriptstyle \rm F}0}}
\symdef{\Kzero}{K_0}

\symdef{\lambdabar}{\overline\lambda}
\symdef{\LdotS}{\bbox{\sigma\cdot L}}
\symdef{\lsim}{\lower0.6ex\vbox{\hbox{$\ \buildrel{\textstyle <}
         \over{\sim}\ $}}}
\symdef{\lzero}{l_{0}}

\symdef{\Mbar}{\overline M}
\symdef{\Mbarzero}{\Mbar_0}
\symdef{\MeV}{\mbox{\,MeV}}
\symdef{\momega}{m_\omegav}
\symdef{\mpi}{m_\pi}
\symdef{\mrho}{m_\rho}
\symdef{\ms}{m_{\rm s}}
\symdef{\Mstar}{M^\ast}
\symdef{\Mstarzero}{M^\ast_0}
\symdef{\mv}{m_{\rm v}}
\symdef{\mzero}{{\rm v}_{0}}

\symdef{\Nbar}{\overline N}

\symdef{\omegaV}{V}
\symdef{\omegav}{{\rm v}}

\symdef{\Phizero}{\Phi_0}
\symdef{\psibar}{\overline\psi}
\symdef{\psidagger}{\psi^\dagger}
\symdef{\pvec}{{\bf p}}

\symdef{\rhoB}{\rho_{{\scriptscriptstyle \rm B}}}
\symdef{\rhoBt}{\wt\rho_{{\scriptscriptstyle \rm B}}}
\symdef{\rhoBzero}{\rho_{{\scriptscriptstyle \rm B}0}}
\symdef{\rhominus}{\rho_{-}}
\symdef{\rhoplus}{\rho_{+}}
\symdef{\rhos}{\rho_{{\scriptstyle \rm s}}}
\symdef{\rhospzero}{\rho'_{{\scriptstyle {\rm s} 0}}}
\symdef{\rhost}{\wt\rho_{{\scriptstyle \rm s}}}
\symdef{\rhoszero}{\rho_{{\scriptstyle {\rm s}0}}}
\symdef{\rhotau}{\rho_{\tauvec}}
\symdef{\rhotaut}{\wt\rho_{\tauvec}}
\symdef{\rhothree}{\rho_{3}}
\symdef{\rhothreet}{\wt\rho_{3}}
\symdef{\rhozero}{\rho_0}

\symdef{\scalar}{\rhos}
\symdef{\scalarN}{{\rhost}}
\symdef{\Szero}{S_0}

\symdef{\tauvec}{{\bbox{\tau}}}
\symdef{\tauthree}{\tau_3}
\symdef{\tensor}{{{s}}}
\symdef{\tensorN}{\wt\tensor}
\symdef{\tensort}{\wt{{s}}}
\symdef{\tensorthree}{\tensor_3}
\symdef{\tensorthreet}{\wt\tensorthree}
\symdef{\Tr}{{\rm Tr\,}}

\symdef{\umu}{u^\mu}
\symdef{\Ualpha}{U_{\alpha}}
\symdef{\Ueff}{U_{\rm eff}}
\symdef{\Uzero}{U_0}
\symdef{\Uzerop}{U_0'}
\symdef{\Uzeropp}{U_0''}

\symdef{\vecalpha}{{\bbox{\alpha}}}
\symdef{\veccdot}{{\bbox{\cdot}}}
\symdef{\vecnabla}{{\bbox{\nabla}}}
\symdef{\vecpi}{{\bbox{\pi}}}
\symdef{\vectau}{{\bbox{\tau}}}
\symdef{\vector}{j_{\scriptscriptstyle V}}
\symdef{\vectorN}{{\wt\vector}}
\symdef{\vecx}{{\bf x}}
\symdef{\Vopt}{V_{\rm opt}}
\symdef{\Vzero}{V_0}

\symdef{\wt}{\widetilde}
\symdef{\wzero}{w_0}
\symdef{\Wzero}{W_0}

\symdef{\zetabar}{\overline\zeta}
%
%
%
%

\def\beq{\begin{equation}}
\def\eeq{\end{equation}}
\def\beqa{\begin{eqnarray}}
\def\eeqa{\end{eqnarray}}

\draft

\preprint{IU/NTC\ \ 99--11}

\title{Large Lorentz Scalar and Vector Potentials in Nuclei}

\author{R. J. Furnstahl and Brian D. Serot\thanks{%
        Permanent address: Department of Physics and Nuclear Theory Center,
         Indiana University,\ \ Bloomington, IN\ \ 47405;
		 serot@iucf.indiana.edu.}}

\address{Department of Physics \\
         The Ohio State University,\ \ Columbus, OH\ \ 43210}
%
%
%

%
\date{December, 1999}
\maketitle
\begin{abstract}
In nonrelativistic models of nuclei,
the underlying mass scales of low-energy quantum chromodynamics (QCD)
are largely hidden.
In contrast,
the covariant formulations used in relativistic phenomenology
manifest the QCD scales in nuclei through
large Lorentz scalar
and four-vector nucleon self-energies.
The abundant and varied evidence in support of this connection 
and the consequences are reviewed.
\end{abstract}

\smallskip
\pacs{PACS number(s): 21.30.-x,12.39.Fe,12.38.Lg,24.85+p}

\section{Prolog}

Large Lorentz scalar and four-vector nucleon self-energies or
optical potentials, each several hundred MeV in the interior of
heavy nuclei, are key but controversial ingredients of successful
relativistic phenomenology%
\footnote{In this paper we use ``relativistic phenomenology'' to refer  
only to field-theory based models, and not to the relativistic
hamiltonian models discussed in Ref.~\cite{KEISTER91}.}
\cite{SEROT86,SEROT92}.
The controversy has persisted because there can be no
{\it direct\/} experimental verification (or refutation) of
such large nuclear potentials.
Here we revisit this issue from the modern perspectives
of effective field theory (EFT) and density functional theory 
(DFT) \cite{SEROT97}.
We argue that the large potentials in the
covariant representation used in relativistic phenomenology
are manifestations of the underlying mass scales of low-energy quantum 
chromodynamics (QCD), which are largely hidden in nonrelativistic treatments.

The connection between low-energy QCD scales and nuclear phenomenology can
be made by applying 
Georgi and Manohar's 
Naive Dimensional Analysis (NDA) and naturalness
\cite{GEORGI84b,GEORGI93b,FURNSTAHL97,FRIAR96b}.
These principles prescribe how to count powers of
the pion decay constant $f_\pi \approx 94\,$MeV
and
a larger mass scale $\Lambda$
in effective lagrangians or energy functionals.
The mass scale $\Lambda$ is associated with the 
new physics beyond the pions:
the non-Goldstone boson masses or the nucleon mass.
The signature of these low-energy QCD scales in the coefficients of a 
relativistic point-coupling model \cite{NIKOLAUS97}
was first pointed out by Friar, Lynn, and Madland \cite{FRIAR96}.
Subsequent analyses have extended and supplemented this idea,
testing it in nonrelativistic mean-field models as well as in different
types of relativistic models \cite{FURNSTAHL97,RUSNAK97,FURNSTAHL97c,PANIC99}.
Estimates of contributions to the energy
functional from individual terms, based on NDA power counting,
are {\em quantitatively\/} consistent with direct, high-quality fits
to bulk nuclear observables \cite{FURNSTAHL99b}.
Naturalness based on NDA scales has proved to be a very robust
concept: nuclei know about these scales!

The EFT perspective, with the freedom to redefine and transform fields,
implies that {\it there are infinitely many  representations
of  low-energy QCD physics\/}.
However, not all are equally efficient or physically transparent.
One of the possible choices is between relativistic and nonrelativistic
formulations.
(In the context of EFT, these can be related by the heavy-baryon
expansion.)
We suggest that the relativistic formulation offers greater insight.

Relativistic phenomenology for nuclei has often been motivated
by the need for relativistic kinematics when extrapolating to extreme
conditions of density, temperature, or momentum transfer.
However, this obscures the issue of relativistic vs.\ 
nonrelativistic approaches for nuclei under ordinary conditions.
The important aspect of relativity in ordinary nuclear systems is
{\it not\/} 
that a nucleon's momentum is comparable to its rest mass, but that 
maintaining covariance allows scalars to be distinguished from the
time components of four vectors.

Despite a long history of criticisms of relativistic 
approaches \cite{BRODSKY84,NEGELE85,BROWN87,WILETS88,ZAHED88,COHEN92},
the use of a relativistic formulation should  not itself be a point
of contention.
The EFT/DFT perspective has largely abrogated
the objections, as we discuss more fully in Ref.~\cite{FURNSTAHL00}.
Furthermore,
recent developments in baryon chiral perturbation theory support the
consistency (and usefulness) of a covariant EFT,
with Dirac nucleon fields in a Lorentz invariant effective 
lagrangian \cite{TANG96,ELLIS98,BECHER99,ELLIS99}.
A similar framework  underlies relativistic approaches to nuclei.
In the nuclear medium, a covariant treatment implies distinct
scalar and four-vector nucleon self-energies.
The relevant question is: what are their mean values?
Relativistic phenomenology suggests several hundred MeV in the center
of a heavy nucleus.

Historically, 
the successes of nonrelativistic nuclear phenomenology 
have been cited to cast doubt on the
relevance of large scalar and vector potentials.
But in a nonrelativistic treatment of nuclei, the distinction
between a potential that transforms like a scalar and one that transforms
like the time component of a four-vector is lost.
Because the leading-order contributions of these two types are opposite
in sign,
an underlying large scale characterizing individual covariant potentials
would be hidden in the nonrelativistic central potential.
Furthermore, the EFT expansion implies that
even potentials as large as 300 to 400\,MeV are sufficiently smaller than
the nucleon mass that a nonrelativistic expansion should converge,
if not necessarily optimally.
{\em 
Thus the success of nonrelativistic nuclear phenomenology provides little
direct evidence about covariant potentials.\/}

If there were an approximate
{\it symmetry\/} that enforced the cancellation between
scalar and vector contributions, then it would be desirable to
build the cancellation into any EFT lagrangian or energy functional.
(Chiral symmetry alone does {\it not\/} lead to scalar-vector fine tuning.)
However, if the cancellation is accidental or of 
unknown origin \cite{WEINBERG91}, then
hiding the underlying scales may be counterproductive.
We argue that nuclei fall into the second category, with the relevant
scales set not by the nonrelativistic binding energy and central potential
(tens of MeV), but by the large covariant potentials (hundreds of MeV).
The signals of large underlying scales would  be patterns in the data
that are simply and efficiently explained by large potentials, but 
which require 
more complicated explanations in a nonrelativistic treatment.

Scattered through the literature over many years is evidence to
support our contention that a representation with large fields, which is 
achieved only with a covariant formulation,
is natural.
We believe that in light of the EFT and DFT reinterpretation of
relativistic phenomenology, it is appropriate at this time to compile
and update the arguments to highlight their strengths and weaknesses.
In the following section, 
we give a concise list of 
empirical and theoretical evidence that
large scales are natural for nuclei, with short descriptions
and pointers to more detailed discussions.
We also include with each item
a brief discussion (which we will
call a ``loophole'') of how  large fields could be avoided,
even in a covariant formulation.

\section{QCD Scales in Nuclei}

\begin{enumerate}


\item {\bf Covariant density functionals fit to nuclei.\/}
Conventional density functional theory (DFT)
is based on energy functionals of the ground-state
density of a many-body system, whose extremization yields a variety of
ground-state properties \cite{DREIZLER90}.
In a covariant generalization of DFT applied to nuclei, 
these become functionals
of the ground-state scalar density $\rhos$ as well as the
baryon current $B_\mu$. 
Relativistic mean-field models are analogs of the Kohn--Sham 
formalism of DFT \cite{KOHN65},
with local
scalar and vector fields $\Phi({\bf x})$ and $W({\bf x})$
appearing in the role of relativistic
Kohn--Sham potentials \cite{SEROT97}.
The mean-field models approximate the exact functional, which includes
all higher-order correlations, using powers
of auxiliary meson fields or nucleon densities.

The scalar and vector potentials are determined by extremizing the energy
functional,
which gives rise to a Dirac single-particle hamiltonian.
The isoscalar part (for spherical nuclei) is
\begin{equation}
   h_0 = -i\bbox{\nabla\cdot\alpha} + 
        \beta \Bigl(M-\Phi(r)\Bigr)
      +  W(r)  \ , \label{eq:hdirac}
\end{equation}
where $M$ is the nucleon mass
and we define $\Mstar \equiv M - \Phi$.
It is not necessary to assume that $\Phi$ is simply proportional
to a scalar meson field $\phi$.  
In fact, $\Phi$ could be proportional to $\phi$
(as in conventional quantum hadrodynamic models \cite{SEROT86,SEROT92}), 
or could be expressed as a sum of scalar and
vector densities (as in relativistic point-couplings models \cite{NIKOLAUS97}),
or could be
a nonlinear function of $\phi$ 
(e.g., see Refs.~\cite{ZIMANYI90,SAITO96,BIRO97}).

\begin{figure}[t]
\begin{center}
  \epsfxsize=3.9in
  \epsffile{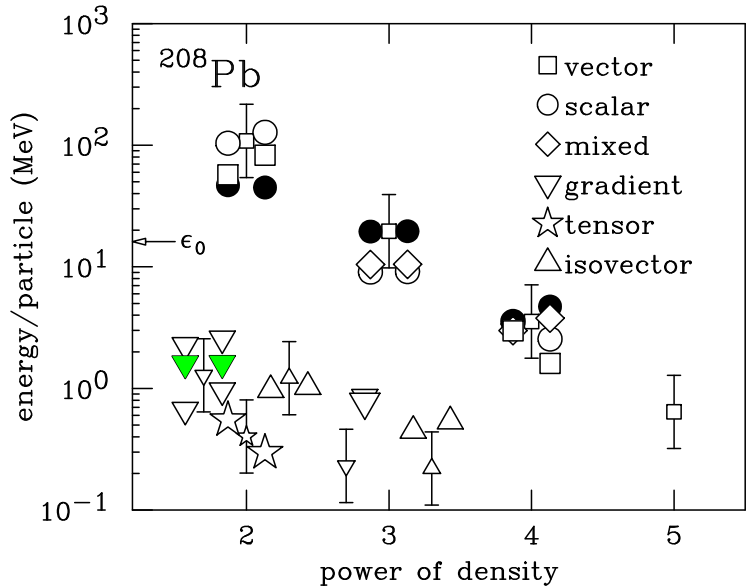}
\end{center}
\caption{Contributions to the energy per particle in ${}^{208}$Pb
 for two relativistic point-coupling models from Ref.~\protect\cite{RUSNAK97}
 are shown as large unfilled symbols (one model on each side of the
 error bars).
 Absolute values are shown.
 The filled symbols denote the sum of the values for each power of the density. 
The small symbols indicate estimates based on NDA, with the error bars
corresponding to natural coefficients from 1/2 to 2.
The binding energy per nucleon in nuclear matter is denoted with $\epsilon_0$.
See Ref.~\protect\cite{FURNSTAHL99b} for more details.}
\label{fig:pb208energy}
\end{figure}

The parameters of the density functional for generalized models
have been determined by detailed
fits to a set of
nuclear properties that should be accurately reproduced according to DFT
\cite{FURNSTAHL97,RUSNAK97}.
Except when a large isoscalar tensor coupling is included,
the scalar and vector potentials $\Phi$ and $W$ are always larger than
300\,MeV in the interior of a nucleus.
These potentials produce a hierarchy of energy contributions
that follow the NDA predictions,
as illustrated by the large unfilled symbols
in Fig.~\ref{fig:pb208energy}.
This agreement persists when correlation corrections are included
explicitly \cite{HU99},
and provides strong evidence that nuclear observables demand
naturally sized parameters.

{\it Loophole: }
     The addition of an isoscalar tensor coupling in the energy
	 functional allows excellent fits to nuclear properties while reducing
	 the size of the scalar and vector potentials slightly,   
     to roughly 250\,MeV \cite{RUFA88,FURNSTAHL98,HUA99}.
	 It should also be noted that
     relativistic formulations of DFT at present lack the rigor of
	 conventional DFT \cite{DREIZLER90,DREIZLER00};  a
	 re-examination from the EFT perspective may improve the situation.


\item {\bf Natural size of leading contribution to binding energy
           per nucleon.\/}
Coefficients in successful relativistic mean-field models are consistent
with naive dimensional analysis (NDA) and naturalness, as expected
in low-energy effective field theories of QCD
\cite{SEROT97,FURNSTAHL97,RUSNAK97}.
If one applies naturalness arguments to the terms in a relativistic
energy functional for nuclear matter and nuclei, the leading scalar and
vector terms at equilibrium density $\rhoB^0$
are each predicted to be of order 
$\rhoB^0/f_\pi^2 \approx 150\,$MeV \cite{FURNSTAHL99b}, independent
of $\Lambda$.
(The $n=2$ energy estimate in Fig.~\ref{fig:pb208energy} is lower 
because it uses an average density in $^{208}$Pb rather than the
peak density in the interior.)
The scalar and vector
potentials $\Phi$ and $W$ in Eq.~(\ref{eq:hdirac}) are each twice
the corresponding energy contributions \cite{RUSNAK97}, 
so they are predicted to be
roughly 300\,MeV in the center of a nucleus. 

{\it Loophole: } Naturalness may give only order-of-magnitude
estimates and there are numerical factors (e.g., combinatoric factors)
that may not be correctly accounted for.
On the other hand, the estimates of contributions to the binding energy, 
recently extended to all terms
in the energy functional, appear to be quite robust \cite{FURNSTAHL99b}
(see Fig.~\ref{fig:pb208energy}).


\item {\bf One-boson-exchange potentials.\/}
The nucleon-nucleon (NN) scattering matrix can be calculated by
unitarizing a kernel for the NN force.  The Lorentz structure of
the kernel follows from covariance, without mentioning degrees
of freedom, but it can be efficiently characterized in terms of
boson exchanges in different physical channels.  This is a very
natural procedure from the point of view of dispersion 
theory \cite{SCOTTI65}.

Each channel is characterized by strength and range parameters
and a cutoff.  A physical interpretation of each is not needed since
the states are virtual in NN scattering.
The parameters are directly related to prominent
resonances in some channels (vector), but not in others (scalar).
Every accurate fit of the parameters to NN observables using
the most general (covariant) kernel has led to an interaction
with large, isoscalar, scalar and vector contributions of comparable
magnitude, but opposite in sign \cite{MACHLEIDT89}.
In the nuclear medium,
these scalar and vector NN amplitudes translate into strong
single-particle potentials, of order several hundred MeV 
at equilibrium density.
These strong potentials persist when short-range correlations
are included explicitly \cite{MACHLEIDT89}.

{\it Loophole: } There may be alternative (covariant) decompositions
 of the kernel that do not result in large scalar and vector
 components (but we are unaware of any!).


\item {\bf Nuclear saturation and observed spin-orbit splittings.\/}
If one adopts a covariant formulation of the energy functional,
the  Lorentz transformation properties of a scalar  component 
induce a velocity dependence in the interaction.
When the functional is fit to nuclear saturation in
nuclear matter, one {\em automatically\/} produces a
spin-orbit force and its observable
consequences in a finite system  
(e.g., nuclear shell structure) \cite{SEROT86,SEROT97}.
Furthermore, the strength of the spin-orbit interaction with 
natural-sized scalar and vector potentials
agrees with the empirical strength
(see Fig.~\ref{fig:oxy} with $\Mstar_0/M\approx 0.60$).
In contrast, the spin-orbit contribution in nonrelativistic energy
functionals must be  adjusted by hand \cite{RING80}.

To our knowledge, there are no simple  alternative explanations
for the origin of the full  spin-orbit strength.
Negele and Vautherin \cite{NEGELE70,NEGELE72} 
tried to take Brueckner calculations of light
nuclei and extract the spin-orbit force from the splittings, but 
found only a fraction of the empirical magnitude,
equal to the result obtained by applying Thomas precession
to the nonrelativistic central potential.
The most sophisticated modern calculations get only {\it half\/} of
the empirical spin-orbit splittings in light nuclei
without including three-nucleon 
forces, and only two-thirds of the splittings using current
three-nucleon-interaction models \cite{PIEPER93,CARLSON99,CARLSONpr}.

{\it Loophole: } An isoscalar tensor term can be used to partially reduce
the scalar and vector potentials while maintaining a large 
spin-orbit splitting (the filled symbols in Fig.~\ref{fig:oxy})
\cite{FURNSTAHL98}.

\begin{figure}[t]
 \setlength{\epsfxsize}{4.1in}
  \centerline{\epsffile{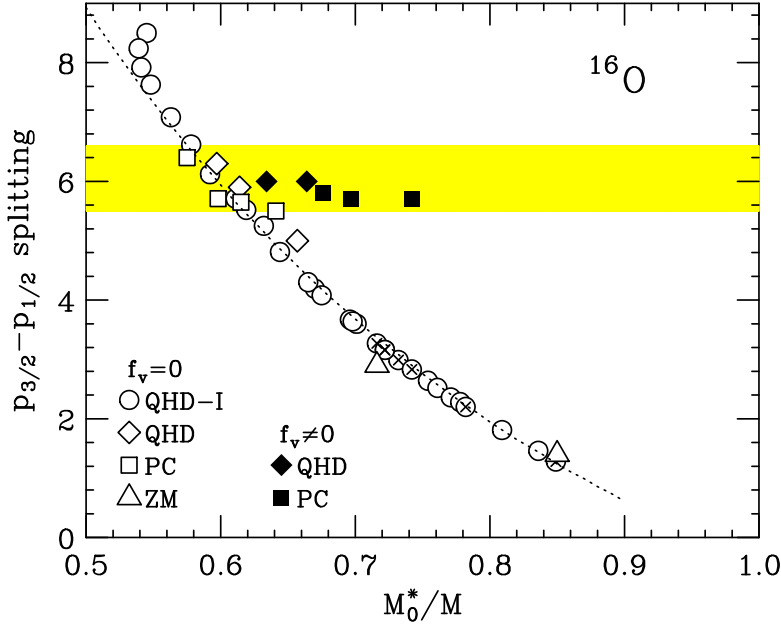}}
\vspace{.1in} 
\caption{\small Spin-orbit splitting for the proton $p$ states in
${}^{16}$O vs.\ equilibrium effective mass $\Mstarzero/M$ 
for a variety of models. 
The shaded band is an estimate of the experimental uncertainty.
Open symbols are for models with isoscalar tensor coupling
$\fv=0$ and filled symbols are for
models with $\fv\neq 0$.
The dotted curve follows from a local density
estimate \protect\cite{FURNSTAHL98}.
See Ref.~\protect\cite{FURNSTAHL98} for descriptions of the models. }
 \label{fig:oxy}
\end{figure}

\begin{figure}[t]
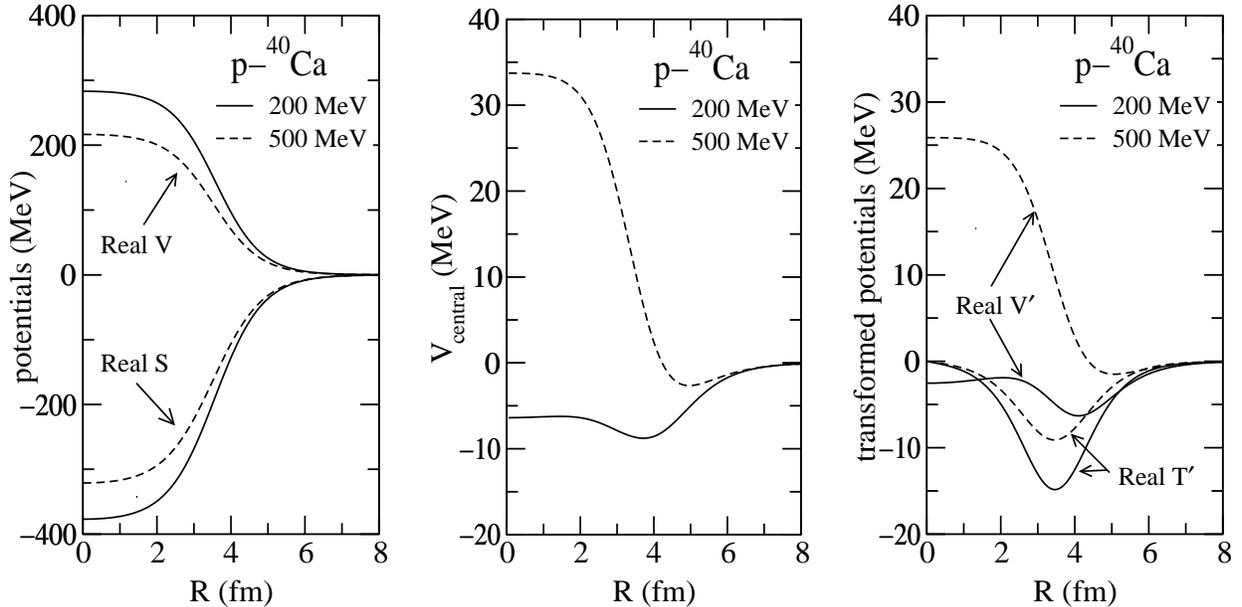

\begin{center}
 \epsfxsize=2.0in
  \epsffile{fig_sv_ca40.eps}
 \hspace*{.10in}
 \epsfxsize=2.0in
 \epsffile{fig_cent_ca40.eps}
 \hspace*{.10in}
 \epsfxsize=2.0in
 \epsffile{fig_vt_ca40.eps}
\end{center}
\caption{The real parts of the scalar ($S$) and vector ($V$) optical
potentials and the corresponding  
nonrelativistic central and transformed vector-tensor
($V'$, $T'$) potentials for proton scattering from 
$^{40}$Ca at two energies.
[The sign convention for the scalar potential $S$ is opposite
to that of $\Phi$ from Eq.~(\protect\ref{eq:hdirac}).] 
The optical potentials are derived from a global fit 
over a wide range in energy \protect\cite{HAMA90,COOPER93}.}
\label{fig:potentials}
\end{figure}


\item {\bf Proton--nucleus scattering spin observables.\/}
In impulse approximation calculations of medium-energy proton--nucleus
scattering spin observables, relativistic treatments with large
scalar and vector optical potentials accurately reproduce the data, while
nonrelativistic treatments are deficient \cite{MCNEIL83,SHEPARD83,CLARK83}.
To get agreement at a similar level in a nonrelativistic formulation,
one has to go beyond the simplest impulse approximation
to include a full-folding treatment \cite{ARELLANO89} 
and medium effects \cite{RIKUS84} (e.g., with a $G$-matrix interaction).
Furthermore,
while the radial shapes of the scalar and vector potentials simply
look like 
the nuclear (baryon) densities with an energy-dependent overall scale,
 the geometries of the nonrelativistic 
optical potentials are much less intuitive and change qualitatively
with different incident energies (see Fig.~\ref{fig:potentials}).
Thus the treatment is clean when natural scales are manifest but
becomes more complicated when they are hidden.

{\it Loophole:  } Large scalar and vector optical potentials can
be transformed away in favor of smaller potentials of different
Lorentz structure \cite{CLARK85}.  
However, the transformed potentials no longer have simple 
radial shapes  
(see Fig.~\ref{fig:potentials}) \cite{CLARK85}.


\item {\bf Energy dependence of the nucleon-nucleus optical potential.\/}
The real part of the empirical optical potential for nucleon--nucleus
scattering up to 100\,MeV incident kinetic energy $\epsilon$
has a well-determined,
nearly linear
energy dependence of $-0.3\epsilon$ \cite{SIEMENS87}.
This energy dependence is directly predicted in a
relativistic mean-field
formulation to be $-(W/M)\epsilon$ \cite{SEROT86},
which is quantitatively correct for a vector potential of natural size.
In contrast, the energy dependence in conventional
nonrelativistic treatments
arises from the non-locality of the exchange corrections in a Hartree--Fock
or Brueckner--Hartree--Fock approximation to the mean-field part of the
optical potential. 
Explicit studies of relativistic calculations
at different approximation levels 
show that the energy dependence is dominated
by the direct contribution
and that exchange corrections are small \cite{KLEINMANN94,TRASOBARES98}.

{\it Loophole: } A direct connection between energy dependence from the
Lorentz structure of the interaction in relativistic formulations and from
exchange corrections in nonrelativistic formulations has not been demonstrated.


\item {\bf Pseudo-spin symmetry.\/}
There is an observed near-degeneracy among sets of energy levels in
medium and heavy nuclei, which have been called {\it pseudo-spin doublets\/}
\cite{HECHT69}.
This degeneracy relies on having a specific relationship between
the nonrelativistic central and spin-orbit potentials.
Ginocchio has shown that this relationship follows
from an $SU(2)$ symmetry of a covariant single-particle hamiltonian if
the nucleon scalar and vector potentials are equal in magnitude
\cite{GINOCCHIO99,GINOCCHIO99b}.
Such a hamiltonian with covariant Kohn--Sham potentials $\Phi$ and $W$
results from the extremization of relativistic energy functionals.

In the exact symmetry limit, with $\Phi = -W$, there are 
no bound positive-energy states, so nuclei do not exist.
However, an approximate pseudo-spin symmetry leading to approximate
pseudo-spin doublets exists for 
$\Phi\approx -W$ \cite{GINOCCHIO99}.
Each potential must be individually large, since their (near) cancellation
must leave a sufficient residual central potential for nuclear binding.

{\it Loophole: }
The symmetry is significantly broken for empirical relativistic potentials
and the consequences of 
this breaking are not understood, so the evidence for pseudo-spin
symmetry is not entirely convincing.
The observed doublets could be accidental or have an unrelated 
origin \cite{BLOKHIN95}.


\item {\bf Correlated two-pion exchange, chiral symmetry, and
          scalar strength.\/}
The scalar-isoscalar part of the NN kernel below 1\,GeV can be
studied in an essentially 
model-independent way in terms of $\pi$--$\pi$
scattering in this channel \cite{JACKSON75,DURSO77,DURSO80}.
Chiral symmetry and unitarity constrain the threshold behavior.
The natural strength of the $\pi$--$\pi$ interaction implies
that the amplitude increases from zero as fast as the unitarity bound.
The predicted integrated strength is consistent with a large
scalar potential \cite{LIN89,LIN90}.
Thus we can understand the origin of the large scale in the scalar
channel from QCD symmetry
constraints, unitarity,  and naturalness.

{\it Loophole: } We do not know of a loophole here.


\item {\bf Cancellations and fine-tuning of nuclear matter.\/}
The small binding energy of nuclear matter would appear to be a
counter argument to the claim that the natural scale for nuclei
is several hundred MeV. 
It would be valid, however, only if nuclear matter were an ordinary,
nonrelativistic Fermi system.
The existence of a minimum in the energy per particle suggests
that different orders in the 
expansion of the energy in powers of the Fermi momentum $\kfermi$
must be comparable.
A logical conclusion is that this should occur only near the underlying
mass scale, where all terms contribute roughly equally, and that
the binding energy should be roughly of this scale.
Yet
the empirical equilibrium conditions are not consistent with this conclusion.

In fact, nuclear matter appears to be an exceptionally fine-tuned 
system, with an equilibrium density far lower than expected
for an ordinary Fermi liquid \cite{JACKSON92,JACKSON94}.  
Covariant formulations offer a compelling explanation:  there {\it is\/} 
an interplay
of different orders in the energy expansion,
but it is highly restricted.  In particular,
repulsion from
the $\kfermi^5$ and $\kfermi^6$ terms becomes important compared
to the attraction from the $\kfermi^3$ piece well below an 
underlying scale of order several hundred MeV.
Furthermore, this happens because the coefficient of the $\kfermi^3$ term
is ``unnaturally'' small,
roughly half the size one would expect from NDA estimates.
In covariant models, this is a direct result of cancellations between
Lorentz scalar and vector contributions that are each of natural size.
The cancellations leading to a small $\kfermi^3$ term do not recur in higher
orders.

This scenario can be tested using $\kfermi$ expansions for the energy
from phenomenologically successful nonrelativistic and relativistic
mean-field models.
If the coefficient of the $\kfermi^3$ term in one of these expansions
is doubled or tripled in size, equilibrium does occur
at much higher density  and the system is bound by 120 to 300~MeV or more
(see Fig.~\ref{fig:sat_double}).
No other term in the expansion exhibits such a sensitivity.
In nonrelativistic models, the cancellation at $\kfermi^3$ has no direct
explanation.
If imposing the cancellation were desirable, one would expect
cancellations to occur at higher orders in the expansion.
However,
an analysis of nonrelativistic 
Skyrme energy functionals finds energy contributions that 
are consistent with NDA counting and a hidden scale at 
leading order only (comparable to the filled symbols at $n=2$ in
Fig.~\ref{fig:pb208energy}) \cite{FURNSTAHL97c}.
Furthermore, we know of no alternative dimensional analysis based
on binding-energy scales that can account for the size of these energy
contributions.

{\it Loophole: } The cancellation in the leading term is only at the
50\% level, which could still be considered natural, without resorting
to explanations based on scalar--vector fine tuning.
In any case, based on the arguments of Jackson \cite{JACKSON92}, 
the extremely low
empirical equilibrium density of nuclear matter is not consistent with a
typical, nonrelativistic, velocity-independent NN potential.

\begin{figure}[t]
\begin{center}
 \epsfxsize=4.1in
 \epsffile{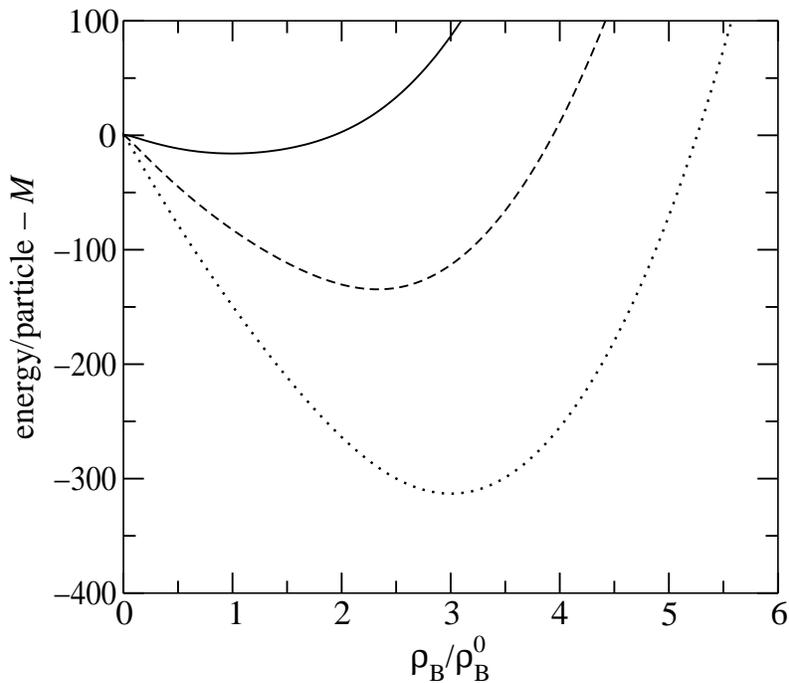}
\end{center}
\caption{Energy per particle in nuclear matter versus density
(normalized to equilibrium density $\rhoB^0$) in a typical
mean-field model fit to nuclei (solid) \protect\cite{RUSNAK97}, 
and  the binding energy curves resulting when the coefficient
of the $\kfermi^3$ term is doubled (dashes) and tripled (dots). }
\label{fig:sat_double}
\end{figure}


\item {\bf Ambiguity in nuclear matter saturation.\/}
If different nonrelativistic potentials, each fit to NN scattering,
are used to calculate the equilibrium binding energy of nuclear matter,
the results fall along a line (the ``Coester line'') with a spread
of 15\,MeV or more.
This spread, which has usually been attributed to ``off-shell'' 
variations in the NN potentials, actually arises because the nuclear matter
calculation requires an interaction that is {\em also} calibrated to three-body
(and, in principle, many-body) {\em on-shell} amplitudes.  This has not
been done in calculations leading to the Coester line, since only two-body
data was used as input in these nonrelativistic calculations.
The magnitude of the variation in equilibrium binding energies
would be difficult to understand if the underlying scale of the two-body
interaction were only 50\,MeV.

Large covariant two-body
potentials in a relativistic formulation, however,
imply sizable 
three-body contributions in 
the corresponding nonrelativistic calculation\cite{WASSON91,SEROT97}.
These contributions are consistent with the spread of the Coester line.
One would expect a smaller spread in relativistic-model
predictions
of the equilibrium point, and this is observed
in practice \cite{MACHLEIDT89,BROCKMANN90}.
Moreover, two-hole-line Dirac--Brueckner--Hartree--Fock calculations
using potentials with natural scales can reproduce both NN scattering
observables and the nuclear matter equilibrium point simultaneously 
\cite{BROCKMANN90}.

{\it Loophole: }  There have been  fewer systematic studies
of relativistic predictions compared to nonrelativistic predictions,
so the relativistic spread may be underestimated.


\item {\bf QCD sum rules for nucleons at finite density.\/}
The QCD sum-rule method relates ground-state matrix elements of QCD operators,
such as the quark condensate, to spectral properties of hadrons
(e.g., masses) \cite{SHIFMAN79,IOFFE81,SHIFMAN92}.
Adapted to finite density, it relates the density dependence of
condensates \cite{DRUKAREV91,FURNSTAHL92a} 
to relative residues at the nucleon quasiparticle poles,
which can be used to predict on-shell scalar and vector self-energies
\cite{COHEN91}.
The mass scales of QCD are directly incorporated into the analysis
through the condensates.

The key feature of the finite-density
sum rule analysis is the {\it covariant\/} decomposition
of a correlator of nucleon interpolating fields.
The quasiparticle pole position is unchanged within the coarse 
resolution of the sum rule approach, but the self-energies extracted
from the correlator residues are sizable. 
This is consistent with weak binding but large covariant potentials. 
A detailed sum-rule
analysis predicts in-medium scalar and vector self-energies
of close to 300\,MeV
(albeit with large error bars) \cite{FURNSTAHL92,COHEN95,FURNSTAHL96c}.

{\it Loophole: } Many provisional assumptions must be made to carry
out the QCD sum-rule analysis \cite{LEINWEBER97}.


\end{enumerate}

\section{Epilog}

In summary, we have argued that a covariant formulation of nuclear
physics has the advantage of manifesting the underlying scales of QCD.
The common signatures of these scales are large nucleon scalar
and vector self-energies.
This connection is shown through both theoretical and empirical
considerations of naturalness in covariant analyses of NN scattering
and nuclear properties.
The manifestation of scales
translates in many instances into simpler, more efficient, or more natural 
explanations of nuclear phenomena than in nonrelativistic formulations.
Examples include the spin-orbit force, the nucleon--nucleon potential,
the energy dependence of the proton--nucleus optical potential, pseudo-spin
doublets, and the cancellations observed in energy functionals of nuclear
matter.
The pieces of evidence supporting a representation with 
large nucleon scalar and vector potentials,
while not definitive when considered individually,
collectively comprise a compelling positive argument.

The evidence shows that the natural scales are not introduced
``artificially'' in a covariant formalism, and that the small binding energy
(2\%) of nuclear matter
arises because it is a finely tuned fermionic system.

Of course, the argument would be moot if there were
direct experimental evidence that rules out the possibility
of large potentials.
We are unaware of any such evidence.
At one time it was thought that predictions in relativistic models for 
isoscalar magnetic moments of odd-$A$ nuclei
are strongly enhanced compared to the data, 
which are close to the Schmidt predictions \cite{MILLER75}. 
Naively, the baryon current of a nucleus with a valence nucleon of momentum
$p$ 
outside a closed shell is $p/\Mstar$, compared to the Schmidt current
$p/M$.
However, if the calculation is forced to respect Lorentz covariance
and the first law of 
thermodynamics, the nuclear current is constrained to be $p/\mu$,
where $\mu \approx M$ is the chemical potential 
\cite{FURNSTAHL91}.
Thus there is no enhancement in a consistent relativistic framework.
The situation for currents at low $q>0$ is still an open problem,
and should be re-examined
in the context of modern EFT-inspired models.

We have emphasized throughout that relativistic and nonrelativistic
formulations are not mutually exclusive alternatives: both should work,
although possibly in very different ways.
Parallel EFT model calculations of the phenomena discussed here,
such as the energy dependence of the optical potential, would more
firmly establish the connections between relativistic and nonrelativistic
explanations.
While we have focused on the naturalness of covariant models, 
 there is also the pragmatic question of the {\it convergence\/}
of relativistic and nonrelativistic EFT expansions.
Large nucleon fields can mean that a Foldy--Wouthuysen  reduction  may
converge slowly, even though $p/M$ is small, because this is also
an expansion in the ratio of potential strength to the nucleon mass.
The relative convergence rates, particularly for spin properties,
merit further examination. 

\vspace*{-2pt}

\acknowledgments

We thank H.~Mueller and N.~Tirfessa for useful comments.
One of us (B.D.S.) thanks the Ohio State University physics department
for its hospitality and financial support during the course of
this research.
This work was supported in part by the National Science Foundation
under Grant No.\  PHY--9800964 and by the U.S. Department of Energy
under Contract No.~DE-FG02-87ER40365.

\end{document}